\documentclass[10pt]{ietbook}
\usepackage[hidelinks,colorlinks=true,citecolor=blue]{hyperref}
\usepackage{graphicx}
\usepackage{array}
\usepackage{booktabs}
\usepackage{authblk}
\usepackage{multirow}
\usepackage{hyperref}
\usepackage{rotating}
\usepackage{cite}

\begin{document}
\rhbooktitle{\textcolor{green}{}}

\markboth{\textcolor{black}{Digital twins for 6G}}{AI-based traffic analysis in DTNs}

\cauthor{Sarah Al-Shareeda\thanks[1]{Artificial Intelligence and Data Engineering Department, Istanbul Technical University, Turkey, email: alshareeda@itu.edu.tr}, Khayal Huseynov\thanks[2]{BTS Group, Istanbul, Turkey, email: khayal.huseynov@btsgrp.com}, Lal Verda Cakir\thanks[3\label{hi}]{School of Computing, Engineering, and The Built Environment, Edinburgh Napier University, United Kingdom, email: cakirl18@itu.edu.tr, \{c.thomson3, b.canberk\}@napier.ac.uk}, Craig Thomson\textsuperscript{\href{hi}{\textcolor{red}{3}}}, Mehmet Ozdem\thanks[4] {Turk Telekom, Istanbul, Turkey, email: mehmet.ozdem@turktelekom.com.tr}, and Berk Canberk\textsuperscript{\href{hi}{\textcolor{red}{3}}}}

\setcounter{chapter}{3}
\chapter{AI-based traffic analysis in digital twin networks}\label{ai-intro}

In today's networked world, Digital Twin Networks (DTNs) are revolutionizing how we understand and optimize physical networks. These networks, also known as 'Digital Twin Networks (DTNs)' or 'Networks Digital Twins (NDTs),' encompass many physical networks, from cellular and wireless to optical and satellite. They leverage computational power and AI capabilities to provide virtual representations, leading to highly refined recommendations for real-world network challenges. Within DTNs, tasks include network performance enhancement, latency optimization, energy efficiency, and more. To achieve these goals, DTNs utilize AI tools such as Machine Learning (ML), Deep Learning (DL), Reinforcement Learning (RL), Federated Learning (FL), and graph-based approaches. However, data quality, scalability, interpretability, and security challenges necessitate strategies prioritizing transparency, fairness, privacy, and accountability. This chapter delves into the world of AI-driven traffic analysis within DTNs. It explores DTNs' development efforts, tasks, AI models, and challenges while offering insights into how AI can enhance these dynamic networks. Through this journey, readers will gain a deeper understanding of the pivotal role AI plays in the ever-evolving landscape of networked systems.

\section{DTNs ecosystem}
\begin{sidewaysfigure}[!htbp]
    \centering
    \includegraphics[width=.7\linewidth]{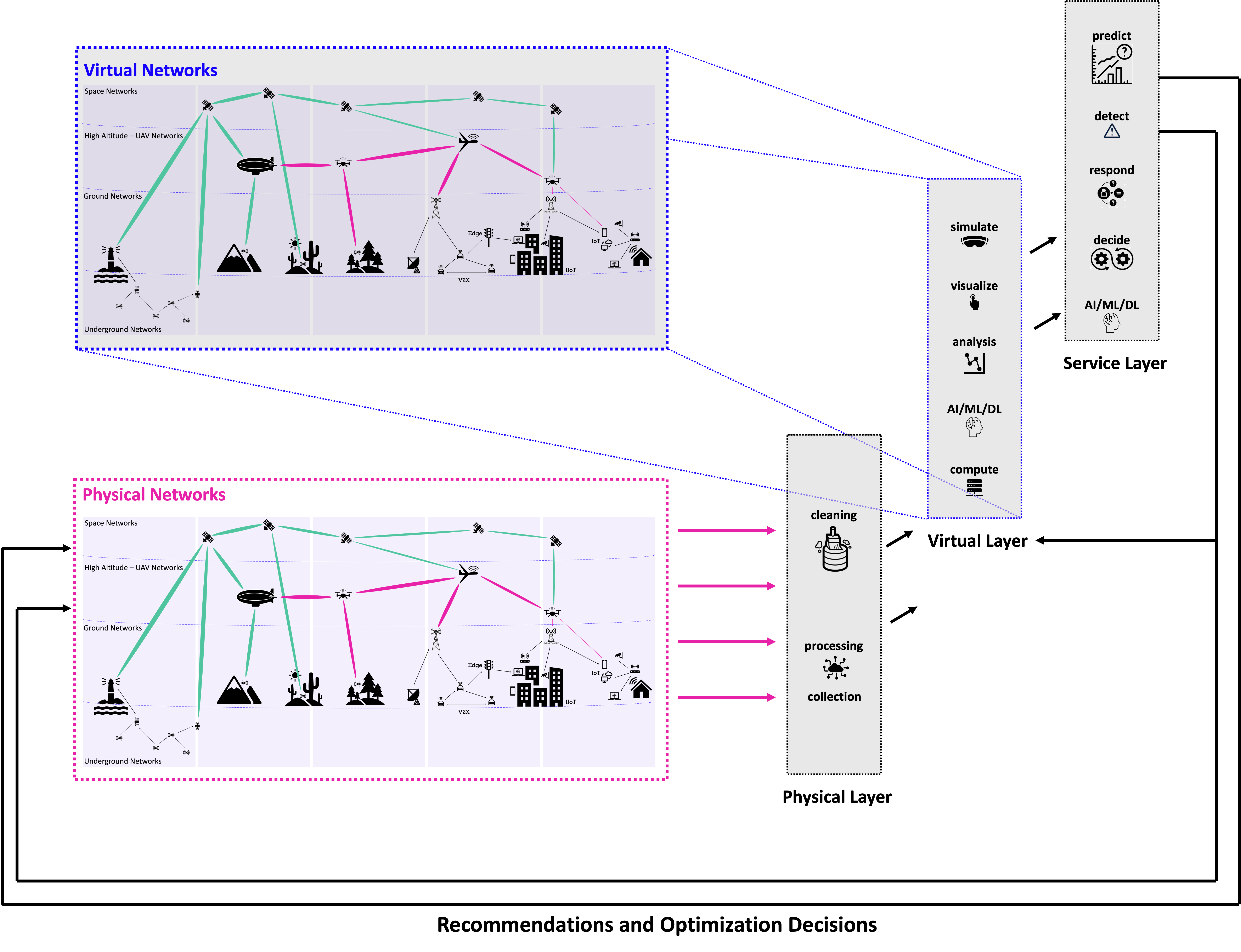}
    \caption{DTNs Descriptive Ecosystem} \label{fig:networktwin}
\end{sidewaysfigure}

DTNs establish a parallel digital domain that mirrors a wide spectrum of physical networks encompassing diverse realms such as wireless networks, mobile networks 4G/5G/5G+/6G, optical networks, underwater, ground, campus, vehicular, airborne, satellite networks, and even space networks. By seamlessly integrating with and harnessing the computational power and AI capabilities intrinsic to this expansive landscape, DTNs transcend conventional boundaries \cite{rathore2021role}. This empowerment allows for the rapid execution of simulations and predictive processes within the mirrored environment, ultimately formulating highly refined feedback recommendations tailored to the difficulties of the associated real-world network.
These refined recommendations, shaped by the marriage of computational power and domain-specific expertise, are then seamlessly disseminated to the actual entities within the physical network. This entire process manifests as a self-sustaining closed-loop data transmission paradigm, where the DTN continuously learns, evolves, and adapts based on the outcomes of its recommendations in the real world.
The architecture of the DTN is partitioned into three layers (sometimes two of these three layers are further split into two sublayers) \cite{tang2022intelligent,huang2022network,sarah1,sarah2}, as depicted in Figure \ref{fig:networktwin}:
\begin{itemize}
    \item Physical Layer: This layer aggregates and preprocesses data from the physical network. The linkage between the physical layer of the DT and the corresponding physical network is referred to as intra-twin communication.
    \item Virtual Layer: The virtual layer of the DT crafts a virtual emulation of the underlying physical network, therein orchestrating the analysis and computation of the collected cleaned data through the integration of AI-driven Machine/Deep Learning (ML)/(DL) capabilities.
    \item Service or Decision Layer: The service or decision layer leverages the insights from the virtual layer to engender well-informed decisions and recommendations for optimizations within the physical realm via intra-twin communication.
\end{itemize}

Intrinsically intertwined within the DTN fabric, virtual entities use inter-twin links to establish intercommunication. The virtual twin architecture capitalizes on simulation, analysis, and optimization capacities to exhibit tangible enhancements across diverse aspects of the physical network, encompassing connectivity, robust communication, security protocols, scalability, Quality of Service (QoS) considerations, routing efficiency, and data-driven decision-making mechanisms. Comprehensive surveys have exhibited DTNs' multifaceted dimensions, potential, and transformative impact across various sectors. Wu et al. \cite{wu2021digital} unveil a panoramic view of the emerging concept of DTNs, delving into their foundational features, definitions, technologies, challenges, and potential application scenarios. A parallel work of Tang et al. \cite{tang2022survey} navigates the convergence of DT technology and 6G wireless systems, probing the realm of DT edge networks. Their investigation involves integrating DT technology with mobile/multi-access edge computing to fortify network performance, security, and cost-effectiveness. Similarly, Kuruvatti et al. \cite{kuruvatti2022empowering} analyze existing literature, tracing the deployment of DT technology in the context of 6G, and engage in discussions regarding use cases, standards development, and the trajectory of future research.
M. Mashaly \cite{mashaly2021connecting} emphasizes the pivotal role of latency, efficiency, and security in effectuating successful DTN implementations. In a quest for novel techniques, McManus et al. \cite{mcmanus2022digital} explore emergent strategies that facilitate data-driven control in new environments, encompassing SLAM-based sensing, network softwarization, reinforcement learning, and collaborative testing techniques that pave the way for robust DTN constructions. Kroyer and Holzinger \cite{kroyer2022digital} dissect DT technology, plumbing its functionalities, core objectives, and technical prerequisites to align capabilities with specific application domains.

Transitioning towards Industry 4.0 integration, Zeb et al. \cite{zeb2022industrial} delve into the relationship between DT technology and Industry 4.0 Internet of Things (IoT) networks. Their survey underscores the role of DTs in optimizing industrial processes, accentuating their interaction with IoT, cloud computing, ML, and advanced analytics. Similarly, Hakiri et al. \cite{hakiri4535810comprehensive} showcase the power of such virtual twin networks to revolutionize industries by interconnecting products, machinery, and human endeavors. Their analysis addresses challenges and underscores how DTs boost maintainability, flexibility, and responsiveness within the Industrial IoT. On the other hand, Su et al. \cite{su2023survey} traverse the evolving landscape of the Internet of DTs, scrutinizing its architecture, technologies, security, and privacy concerns. Concurrently, Khan et al. \cite{khan2023federated} survey the application of Federated Learning (FL) in DT-enabled vehicular networks, exploring the fusion of FL and DT modeling to meet diverse latency, reliability, and experiential quality requirements. In addition to the survey above studies, numerous endeavors have been put forth to establish and fabricate DTs for a diverse array of network domains, encompassing cellular networks, wireless networks, optical networks, satellite and aeronautic networks, vehicular networks, and industrial IoT networks. The subsequent section provides an extensive overview of these concerted initiatives in constructing DTNs.

\section{DTNs development efforts: literature review}
\subsection{Networks in general}
Many efforts have been made to develop the DTNs and explore their potential. Among significant contributions, R. Larsson \cite{larsson2021creating} explores network virtualization integration, focusing on scalability using domain-specific languages, providing insights into potential virtualization technologies within the DTN landscape. Addressing the challenges posed by emerging technologies, Hui et al. \cite{hui2022digital} propose DTNs as tools for "What-if" evaluations, addressing emerging technology challenges by leveraging data-driven methods. Similarly, S. Kislyakov \cite{kislyakov2022digital} contributes a theoretical framework for defining DTNs within Industry 4.0 contexts, establishing clear communication models, and determining the roles of various models. Exploring the intricacies of handling diverse network data, Yang et al. \cite{yang2022multi} develop a sophisticated DTN system capable of managing multi-source heterogeneous network data, facilitating a holistic view of network performance. Innovating the architecture of future networks, Tan et al. \cite{tan2021toward} propose a cyber DT-based framework that supports intelligent services and resource management across both cyber and physical domains, heralding a new era of network adaptability. Additionally, Zhu et al. \cite{zhu2021knowledge} introduce a comprehensive DTN architecture, enriching network capabilities through multi-layer integration, creating a unified framework for enhanced connectivity.

Contributing to the deployment methodology of DTs, Luan et al. \cite{luan2021paradigm} highlight the transformative potential of DTNs in leveraging computing capabilities. Hamzaoui and Julien \cite{hamzaoui2022social} present a structured approach that addresses DTs' interactional and systemic aspects, harmonizing the development of DT technology across various fields. On the other hand, Chen et al. \cite{chen2022classification} delve into the classification of DTN models, paving the way for orchestrated simulations and optimized performance verification. Building on this classification framework, Szanto et al. \cite{szanto2022digital} categorize DTs based on application roles, demonstrating the power of bi-directional communication through real-time protocols. These diverse contributions underscore the profound potential of DTNs to revolutionize network paradigms, redefine connectivity, and bridge the gap between physical and digital realms.

\subsection{Cellular networks: 5G and beyond}
Efforts in developing DTNs for mobile networks are increasing, driven by a collective aspiration to enhance management practices and elevate operational efficiency. The investigations within this domain have yielded notable works. In the context of 5G, Tao et al. \cite{tao2021digital} delve into the role of DTNs as a catalyst for advancements, traversing architectural nuances and transformative technologies. Likewise, Seilov et al. \cite{seilov2021concept} meticulously tailor DTNs to intricacies within the telecom landscape, enhancing the lifecycle with streamlined processes and vigilant traffic monitoring. Likewise, Tao et al. \cite{tao2023deep} introduce a thought-provoking paradigm, suggesting the data-driven modeling of 5G core networks through DT frameworks. Other works navigate the intricate terrain of optimizing mobile networks, weaving a synergy between NDTs and AI/ML Jang et al. \cite{jang2023digital, rodrigo2023digital, pantovic20225g}.

In the landscape of 6G, Lu et al. \cite{lu2021adaptive} ingeniously embed DT technology within the edge networks, harnessing AI to orchestrate optimal network efficiency. Lin et al. \cite{lin20236g} pivot towards real-time DTNs, outlining pragmatic implementations alongside AI/ML-driven optimization strategies. The endeavor of Duong et al. \cite{duong2023digital} embarks on a journey through multi-tier computing within the 6G realm, with AI, to boost the metaverse's architectural foundations. Guo et al. \cite{guo2023five} spotlight the application of DT technology in 6G networks, specifically emphasizing heightening the QoS for mobile devices and applications. Masaracchia et al. \cite{masaracchia2023digital} reveal the potential of DT collaboration for fortifying intelligent and resilient Radio Access Networks (RANs) in the 6G landscape.

Further studies encompass diverse explorations, each carving a distinct niche within the expansive landscape of DTNs for mobile networks. Collectively, these endeavors resonate as a testament to the potential of DTNs specifically calibrated to harmonize with the intricacies of mobile networks \cite{sheen2020digital,ahmadi2021networked,shu2022digital,vila2023design,gong2023scalable,mirzaei2023network,raza2023definition, apostolakis2023digital}.

\subsection{Wireless networks}
Efforts to develop DTNs for wireless networks have also been explored, aiming to organize data flows and enhance connectivity. Almeida et al. \cite{almeida2023position} introduce DTs to wireless networks, blending simulation and experimentation. Their position-based ML propagation loss model enhances ns-3 simulations, estimating propagation loss and significantly advancing network evaluation. Bariah et al. \cite{bariah2022digital} envision DTs converging applications, particularly in wireless network domains. Their comprehensive representation of wireless network elements integrates AI for training, reasoning, and decision-making, amplifying wireless technologies in smarter, more sustainable smart cities. These explorations highlight the fusion of DT technologies with modern communication approaches in wireless networks. Further use cases are included in Section \ref{tasks}.

\subsection{Optical networks}
The integration of DTNs into optical networks is rapidly gaining attention, fueled by the capabilities of AI-powered virtual counterparts. Chen et al. \cite{chen2023digital} lead the way by deploying DTNs using ML techniques. This pioneering effort showcases the transformative potential of combining DTNs with AI, reshaping network operations. Q. Zhuge \cite{zhuge2022ai} focuses on DTs for self-driving optical networks, skillfully integrating AI to assess these comprehensive networks' lifecycles comprehensively. This work underscores the ongoing pursuit of adaptable and intelligent optical networks. Kuang et al. \cite{kuang2022construct} delve into the evolution of intelligent optical infrastructures guided by purpose-built algorithms and AI-infused DT models. Likewise, Solmaz et al. \cite{solmaz2022digital} highlight DTs' potential in photonics-based architectures, envisioning energy-efficient smart campuses and emphasizing their diverse capabilities.

Janz et al. \cite{janz2022digital} shed light on AI-empowered optical transmission performance assessment. This entails automating tasks such as provisioning and risk mapping, and enhancing network efficiency through AI interventions. Wang et al. \cite{wang2021role} disrupt the optical communication landscape with AI-driven DTs designed to tackle evolving challenges in efficient system management. Zhuge et al. \cite{zhuge2023building} offers a comprehensive tutorial on AI-driven modeling, telemetry, and self-learning, enhancing DT capabilities to address intricate aspects like optical transmission impairments. Mello et al. \cite{mello2023digital} explore NDTs in data-driven optical networks, focusing on intent-based allocation strategies to improve efficiency and reliability. Authors in \cite{eldeeb2022digital, velasco2023applications} adeptly integrate DT technology with optical wireless communication networks, leveraging AI to enhance reliability and network performance. Their efforts advance misconfiguration detection and Quality of Transmission (QoT) estimation. Similarly, Vilalta et al. \cite{vilalta2022architecture} introduce innovative architectural frameworks for DT optical networks, guiding precise network design and optimization.

\subsection{Satellite and aeronautic networks}
Within non-terrestrial networks such as satellite and aeronautic networks, creating aeronautic DTNs has been presented by Chang et al. \cite{chang2022kid} in 2022. They carve a distinct mark by conceiving a multi-fidelity simulator designed for wireless Unmanned Aerial Vehicle (UAV) networks. In response to the challenges of AI/ML-driven control, their amalgamation of two popular simulators orchestrates synchronized simulations. In parallel, Bilen et al. \cite{bilen2022proof} navigate the intricacies of aeronautical landscapes, enhancing the selection of wireless core networks in dynamically evolving aero-settings as proof of DTN adaptability. Confronting data collection challenges, Moorthy et al. \cite{moorthy2022middleware} steer a multi-fidelity simulator toward wireless UAV networks. The synergy of other simulators ignites a symphony of coordinated simulations, laying the bedrock for DT-infused advancements in UAV applications. The work of Brunelli et al. \cite{brunelli2022framework} tailors a DT model for a 3D urban air mobility interwoven with dynamic links guided by heuristic cost considerations. Zhou et al. \cite{zhou2023hierarchical}, and Al-Hraishawi et al. \cite{al2023digital} lead with developing a hierarchical DTN tailored to the distinctive satellite communication requirements. This endeavor showcases DT technology's ability to overcome challenges, spanning design, emulation, deployment, and maintenance. The hierarchical DTN assumes a dual role, seamlessly intertwining communication and networking twins.

\subsection{Vehicular networks}
In this area of research, there are many efforts to harness DTNs' potential to enhance vehicular networks' performance and efficiency. Palmieri et al. \cite{palmieri2022co} contribute with a method that develops accurate Digital-Twin models for multi-agent vehicular networks. The approach involves simulating components across modeling languages and assessing the impact of network delay using AI techniques. In real-time traffic monitoring, C. Fennell \cite{fennell2022communication} explores Apache Kafka's potential as a communication link between motorway sensors and a DT. The research aims to establish a communication architecture with improved availability, throughput, and low latency, investigating network traffic analysis's impact on latency and throughput. Wang and Chen \cite{wang2023internet} introduce a 5G-based framework for driverless tracked vehicles in the context of the Internet of Vehicles (IoV). Vehicle-to-Everything (V2X) (Vehicle-to-Everything) communication comes to the forefront as Wagner et al. \cite{wagner2023spat} delve into the interaction between traffic light controllers and road vehicles. Investigating state-of-the-art V2X communication technologies, the study develops a traffic control system adhering to V2X protocols. These are a few of the many efforts in this domain; other use cases are highlighted in Section \ref{tasks}.

\subsection{Industrial IoT networks}
Innovative steps in constructing DTNs have been taken for the Industrial IoT (IIoT) networks. Kherbache et al. \cite{kherbache2021digital} propose a comprehensive architecture for IIoT, introducing an NDT tailored for closed-loop network management. Guimaraes et al. \cite{guimaraes1digital} pave the way for automated IoT instrumentation networks through DTs. Isah et al. \cite{isahdata} present a data-driven DTN architecture that bridges the gap between the physical and digital realms. Kherbache et al. \cite{kherbache2022network} design an NDT to enhance IIoT network management and optimization. Jagannath et al. \cite{jagannath2022digital} delve into real-time modeling using DT frameworks within the IoT landscape. Rizwan et al. \cite{rizwan2023intelligent} combine FL and DT technology to empower IoT networks. Hakiri et al. \cite{hakiri2023hyper} embark on developing the Hyper-5G project's NDT, a platform geared towards replicating IoT networks for experimentation, especially in evaluating novel IoT services. These endeavors reshape the fabric of IoT networks, seamlessly integrating the physical and digital dimensions for enhanced connectivity and insightful management.

\section{Key tasks in DTNs analysis} \label{tasks}
\begin{figure}[!t]
  \begin{center}
\centering\includegraphics[width=.7\textwidth]{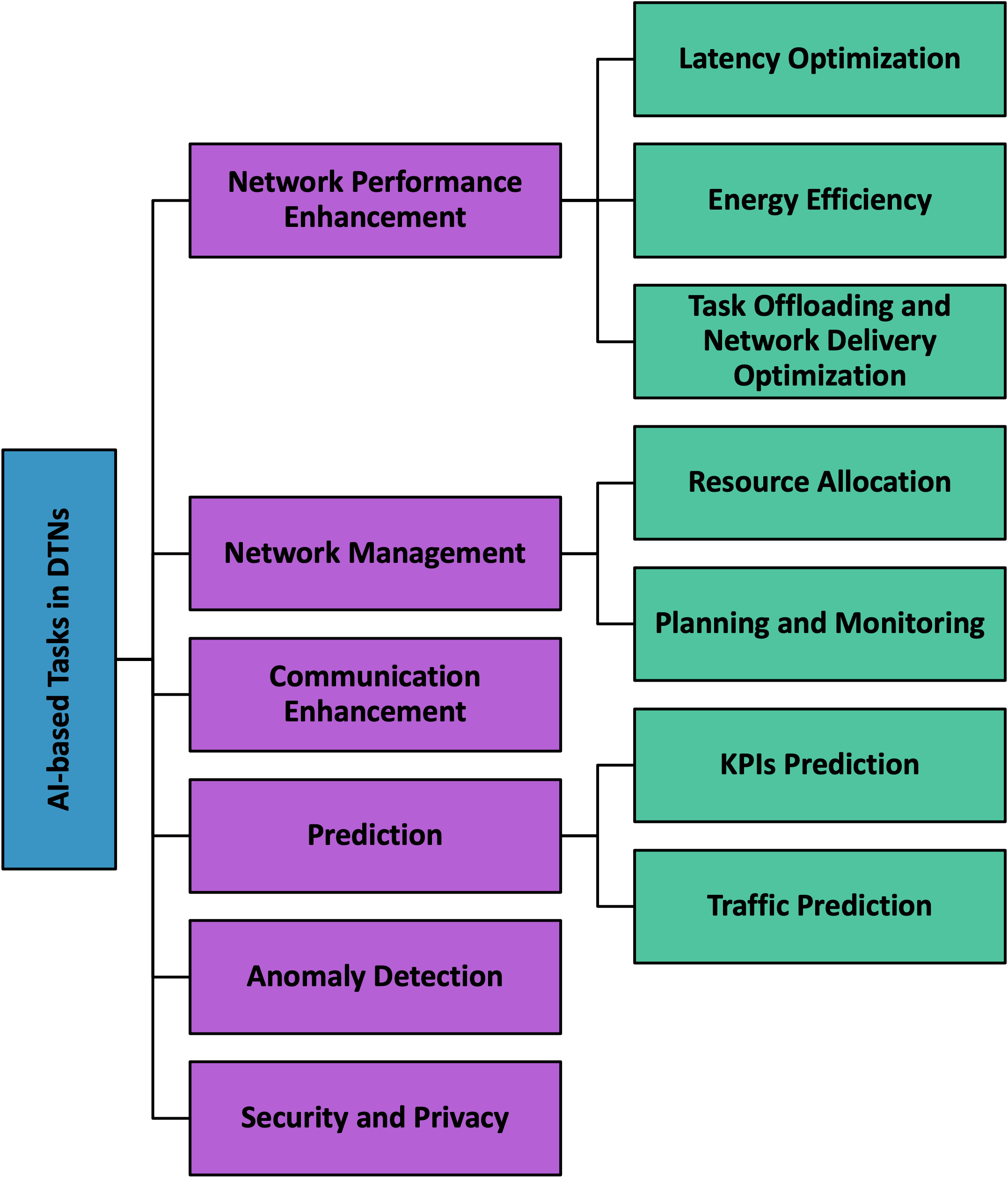}
    \caption{\label{fig:tasks}Key AI-based Tasks in DTNs}
  \end{center}
\end{figure}

The traffic observed within the physical network signifies a complex web of information exchanges and interactions among the tangible components of the network. Following the initial phases of filtering, refining, and preprocessing at the physical layer of the DT, this traffic seamlessly traverses to the virtual layer of the twin architecture. This particular layer provides a versatile foundation that adeptly handles the following various facets, Figure \ref{fig:tasks}, of network enhancement, resource management, communication optimization, predictive analysis, anomaly detection, and security and privacy assurance:
\begin{enumerate}
\item Network Performance Enhancement: A pivotal role of the DTN is optimizing the network's performance. It maintains a constant vigil over data exchanges, ensuring the network operates at its best efficiency. This proactive stance allows for the early identification and swift resolution of emerging performance issues, safeguarding network operations' uninterrupted and efficient flow; Subsection \ref{performanceenhancement} covers the related AI-based works.
\item Network Management: Efficiently overseeing network resources is a keystone of DTN functionality. It encompasses the sound allocation of resources, dynamic load balancing, and adept network configuration to guarantee seamless and reliable network communication; Subsection \ref{networkmanagement} addresses this key task of DTN.
\item Communication Enhancement: DTN tirelessly endeavors to augment communication within the network. It strives to streamline data transmission, minimize latency, and ensure uninterrupted connectivity. This is quintessential for perpetuating efficient data exchange amongst network constituents and users; Subsection \ref{communicationenhancement} unravels the literature and the used AI-based tools and techniques.
\item Prediction Analysis: Armed with historical data, the DTN's predictive capabilities forecast forthcoming network trends and potential disruptions. This foresight empowers preemptive interventions, whether bolstering resources in anticipation of increased demand or proactively addressing impending issues before they impact network performance; Subsection \ref{prediction} covers the related AI-based works.
\item Anomaly Detection: DTN's strength lies in rapidly identifying anomalies and irregularities within the network. Regarding security breaches, technical glitches, or atypical network behavior, DTN should swiftly alert administrators, enabling them to take immediate remedial action; Section \ref{anomaly} covers the related AI-based works.
\item Security and Privacy: Network security and data privacy are non-negotiable. DTN methodically scrutinizes data flows, unearths vulnerabilities, and fortifies network defenses against potential threats. Moreover, it ensures strict adherence to data privacy regulations, safeguarding sensitive information from unauthorized access; Subsection \ref{security} covers the related AI-based works.
\end{enumerate}

Harmoniously executed tasks empower DTN to offer a holistic suite of analytical capabilities, fostering proactive traffic analysis and optimization. Advanced AI-driven models and analytical tools are strategically integrated throughout the DTN architecture to facilitate effectively executing these tasks. This collaborative endeavor culminates in elevated network efficiency and heightened reliability from the initial stages of data aggregation to the culminating decision-making phase. It significantly contributes to the overall optimization of network operations.

\begin{boxes}
{\boxhead{Key Tasks in DTNs Analysis}}
{\begin{itemize}
\item \textbf{Network Performance Enhancement:} Ensuring best network operation.
\item \textbf{Network Management:} Efficiently allocating network resources.
\item \textbf{Communication Enhancement:} Optimizing data exchange and connectivity.
\item \textbf{Prediction Analysis:} Forecasting future network trends and disruptions.
\item \textbf{Anomaly and Fault Detection:} Swiftly identifying irregularities and threats.
\item \textbf{Security and Privacy Analysis:} Protecting network integrity and data privacy.
\end{itemize}}
\end{boxes}

\subsection{AI-Based Network Performance Enhancement}\label{performanceenhancement}
A substantial body of academic research on enhancing network performance through AI exists. This body of research delves into incorporating DTNs to strengthen the effectiveness of different types of networks. These studies collectively examine the confluence of AI, ML, and DL techniques to tackle challenges in network operation and ultimately achieve improved performance. Notably, the research and literature address these primary facets of network performance:

\subsubsection{Latency Optimization}
In the pursuit of optimizing network performance by reducing latency, several research endeavors have emerged. Saravanan et al. \cite{saravanan2021performance} employ innovative approaches grounded in DTs to enhance mobile edge computing within the dynamic 6G network environment. Leveraging AI techniques, specifically the Lyapunov approach and Actor-Critic (A3C) learning, they address the intricate challenges arising from user mobility and the volatile nature of edge computing environments. Yang et al. \cite{yang2021systematic} introduce a flow emulation framework tailored for delay-tolerant networks. They emphasize accurately replicating network traffic patterns, offering insights into optimizing latency. Along the same lines, Van Huynh et al. \cite{van2022fairness} take on the noteworthy challenge of achieving fairness-aware latency minimization in DT-aided edge computing. In \cite{van2022urllc,van2022digital}, the same group of authors concentrates on minimizing latency in computation offloading, particularly within IIoT environments, utilizing DT wireless edge networks. Wang et al. \cite{wang2022optimal} present an innovative Time Sensitive Networking (TSN) scheduling technique applied in delay-tolerant networks; their research centers on optimizing data transmission and scheduling within intricate network structures. Tang et al. \cite{tang2022intelligent} endeavor to attain differentiated Service Level Agreements (SLAs), low latency, and deterministic bandwidth in complex network architectures by deploying DTNs. Likewise, Ferriol-Galmes et al. \cite{ferriol2022building} introduce TwinNet, a cutting-edge Graph Neural Network (GNN)-based model designed to estimate QoS metrics accurately. Meanwhile, Duong et al. \cite{duong2022digital} shift their research focus towards comprehensive end-to-end latency minimization within DT-aided offloading scenarios in UAV networks. Their considerations extend to Ultra-Reliable Low Latency Communication (URLLC) and UAVs. Jiang et al. \cite{jiang2022virtual} employ the Proximal Policy Optimization (PPO) AI approach to optimize resource-intensive task computing in dynamic UAV networks. In parallel, Li et al. \cite{li2022digital} delve into intelligent task offloading within UAV-enabled Mobile Edge Computing (MEC) environments, utilizing DTs. They employ the Double Deep Q-network (DDQN) algorithm for optimization. Moreover, Wang et al. \cite{wang2023digital} delve into the intricate realm of computation offloading decisions for UAVs, particularly in disaster scenarios. Their research leverages the Upper Confidence Bound (UCB)-based stable matching algorithm to enhance decision-making processes, ultimately optimizing latency and network performance.

\begin{table}[!tbhp]
\processtable{Latency Optimization Works}
{\begin{tabular*}{\textwidth}{@{\extracolsep{\fill}}ll@{}}\toprule
\textbf{Work} & \textbf{Utilized AI Tools} \\
\midrule
Saravanan et al. \cite{saravanan2021performance} & Lyapunov approach and A3C learning \\
Yang et al. \cite{yang2021systematic} &emulation framework \\
Van Huynh et al. \cite{van2022fairness, van2022urllc, van2022digital} & N/A \\
Wang et al. \cite{wang2022optimal} & Time Sensitive Scheduling \\
Tang et al. \cite{tang2022intelligent} & N/A \\
Ferriol-Galmes et al. \cite{ferriol2022building} & TwinNet (GNN) \\
Duong et al. \cite{duong2022digital} & N/A \\
Jiang et al. \cite{jiang2022virtual} & PPO \\
Li et al. \cite{li2022digital} & DDQN \\
Wang et al. \cite{wang2023digital} & UCB-based stable matching algorithm \\
\botrule
\end{tabular*}}{}
\end{table}

\subsubsection{Energy Efficiency}
Several noteworthy studies have emerged in the context of network energy efficiency enhancement. Zhao et al. \cite{zhao2022energy} focus on optimizing energy consumption and computational overhead in 6G networks through an innovative DT-edge association scheme. Chen et al. \cite{chen2023a3c} explore the integration of MEC and DTs to minimize energy consumption, utilizing the asynchronous advantage A3C method for optimization. Shui et al. \cite{shui2023cell} tackle reducing energy usage in cell-free networks, maintaining URLLC requirements, and employing AI optimization methods such as DQN and DT. In IoT networks, Liu et al. \cite{liu2023deep} address energy efficiency issues, leveraging AI, ML, and Deep Reinforcement Learning (DRL) to optimize energy consumption through resource prediction and migration strategies. These studies collectively contribute to pursuing more sustainable and efficient network infrastructures.

\begin{table}[!tbhp]
\processtable{Energy Efficiency Enhancement Works}
{\begin{tabular*}{\textwidth}{@{\extracolsep{\fill}}ll@{}}\toprule
\textbf{Work} & \textbf{Used AI Tool} \\
\midrule
Zhao et al. \cite{zhao2022energy} & association scheme \\
Chen et al. \cite{chen2023a3c} & A3C method \\
Shui et al. \cite{shui2023cell} & DQN\\
Liu et al. \cite{liu2023deep} & ML and DRL \\
\botrule
\end{tabular*}}{}
\end{table}

\subsubsection{Task Offloading and Content Delivery Optimization}
In the rapidly evolving realm of the Internet of Everything (IoE), driven by advancements in 5G/6G and AI technologies, Yi et al. \cite{yi2022digital} create a highly efficient content delivery system specially designed for the IoE environment, and they have employed DT technology for this purpose. Their primary goal is to fine-tune content delivery using advanced AI techniques like Long Short-Term Memory (LSTM) and RL. Similarly, Gao et al. \cite{gao2023quality} are tackling the impending challenges posed by the massive content delivery demands expected in the 5G and 6G communication era. Their strategy is to integrate DT technology into edge networks, enabling precise measurement of Quality-of-Decision (QoD) and Quality-of-Experience (QoE) metrics. These metrics, in turn, inform content delivery strategies, catering to both machine-centric and user-centric communication scenarios. Moving forward, Guemes-Palau et al. \cite{guemes2022accelerating} delve into the complexity of network management, where they emphasize the use of DTNs and optimization techniques. Their approach involves harnessing the power of DRL and Evolutionary Strategies (ES) to optimize network performance. In a related endeavor, Lin et al. \cite{lin2022digital} introduce a novel packet-action sequence model, a tool for accurately modeling network behavior within DTNs. Their work focuses on understanding the dynamic nature of network behaviors, highlighting the crucial role of AI in comprehending and optimizing these intricate patterns.

Shifting gears slightly, Ursu et al. \cite{ursu2023towards} have directed their attention toward optimizing cloud network functions, specifically HTTP load balancing. Their research takes place within Kubernetes (K8s) cluster environments. It enhances the modeling and optimization of network function behaviors using DTs and ML techniques. Further up in the skies, Zhao et al. \cite{zhao2022interlink} tackle challenges related to low-orbit satellite networks integrated with terrestrial networks using DTs. They employ genetic algorithm techniques to optimize network performance, reduce handover frequency, enhance data delivery, and improve network efficiency. Sun et al. \cite{sun2020dynamic} aim to elevate air-ground network performance using FL. Their approach involves collaborative model training without the need to share sensitive data. Dynamic DTs are integrated to provide real-time reflections of network status, contributing to this ambitious enhancement effort. Lastly, Han et al. \cite{han2022polymorphic} present a pioneering Polymorphic Learning (PL) framework for DTNs. Their emphasis is on ensuring secure and personalized services. Their contribution lies in optimizing learning algorithms tailored to individualized service needs. Lastly, Liu et al. \cite{liu2021digital} present a pioneering DT-assisted scheme thoughtfully tailored to the emerging landscape of 6G networks. Their efforts revolve around the intelligent offloading of tasks from mobile users to cooperative mobile-edge servers, with the overarching aim of optimizing the utilization of network resources. The intelligent application of AI techniques fortifies this endeavor, including the Decision Tree algorithm and double DL, employed to fine-tune task-offloading decisions, ultimately ushering in enhanced network performance.

\begin{table}[!tbhp]
\processtable{Task Offloading and Content Delivery Optimization Works}
{\begin{tabular*}{\textwidth}{@{\extracolsep{\fill}}ll@{}}\toprule
\textbf{Work} & \textbf{Used AI Tool} \\
\midrule
Yi et al. \cite{yi2022digital} & LSTM, RL \\
Gao et al. \cite{gao2023quality} & QoD, QoE metrics \\
Guemes-Palau et al. \cite{guemes2022accelerating} & DRL, ES \\
Lin et al. \cite{lin2022digital} & Packet-action sequence model\\
Ursu et al. \cite{ursu2023towards} & ML load balancing \\
Zhao et al. \cite{zhao2022interlink} & genetic algorithm \\
Sun et al. \cite{sun2020dynamic} & FL \\
Han et al. \cite{han2022polymorphic} &PL\\
Liu et al. \cite{liu2021digital} & Decision Tree, double DL \\
\botrule
\end{tabular*}}{}
\end{table}

\subsection{AI-Based Network Management}\label{networkmanagement}
\subsubsection{Planning and Monitoring}
Regarding planning and placement, Qu et al. \cite{qu2022general} introduce a Decision Tree framework for intelligent network management, primarily emphasizing network optimization and management strategies. Similarly, Zhao et al. \cite{zhao2022design} propose a network planning system based on DTNs to enhance efficiency through architectural design. Corici and Magedanz \cite{corici2023digital} explore the potential of DTs in optimizing 5G network management. In a related context, Bilen et al. \cite{bilen2023digital} propose a DT framework for managing aeronautical ad-hoc networks within the context of 5G and beyond. Chukhno et al. \cite{chukhno2022placement} focus on network management by dynamically placing social DTs within advanced 5G+/6G networks.
Additionally, they aim to enhance data exchange efficiency and service discovery, improving operational efficiency by optimizing IoT DT placement within the network. In a similar context, Xiao et al. \cite{xiaoevolutive} present an evolutionary framework to optimize server layout within DTNs. In vehicular networks, Zhao et al. \cite{zhao2022elite} propose a DT architecture to enhance vehicular network routing policies, incorporating policy learning and real-time adaptation using DT representations. Fu et al. \cite{fu2023communication} concentrate on improving decision-making in connected autonomous vehicles through a DT-assisted framework, employing hierarchical multi-agent RL to enhance vehicle collaboration and decision-making efficiency. Al-Hamid and Al-Anbuky \cite{al2022vehicular} delve into the modeling and analysis of dynamic groupings in vehicular networks, aiming to assess the performance of dynamic vehicle groups, particularly during the self-healing and self-formation phases.

Monitoring and analysis are critical aspects in the evolving network management landscape. Jiao et al. \cite{jiao2023mobile} explore DT management in tower networking, focusing on system analysis and design methods. Ren et al. \cite{ren2023end} present a comprehensive scheme for managing SLA quality in a two-level Cloud RAN, employing AI techniques such as GraphSAGE, Deep Double Q-Network (DDQN), and Bayesian Convolutional Neural Network (BCNN). Rosello et al. \cite{rosello2023network} introduce the NDT as an experimental and verification framework for 6G technology. Kherbache et al. \cite{kherbache2022digital} emphasize crafting a DTN tailored for the IIoT with a primary goal of real-time intelligent management. Wei et al. \cite{wei2021data} introduce the concept of a DTN to foster innovation in Industry 4.0, specifically highlighting data-driven routing within the DTN framework. Raj et al. \cite{raj2023building} develop a DTN architecture tailored for Software Defined Networking (SDN)-based networks, allowing for monitoring and verification without disrupting the live system. This architecture incorporates Knowledge Graph (KG) and Template-based contextual information.

\begin{table}[!tbhp]
\processtable{Planning and Monitoring Works}
{\begin{tabular*}{\textwidth}{@{\extracolsep{\fill}}ll@{}}\toprule
\textbf{Work} & \textbf{Used AI Tools} \\
\midrule
Qu et al. \cite{qu2022general} & Decision Tree \\
Zhao et al. \cite{zhao2022design} & N/A \\
Corici and Magedanz \cite{corici2023digital} & N/A \\
Bilen et al. \cite{bilen2023digital} & N/A\\
Chukhno et al. \cite{chukhno2022placement} & N/A\\
Xiao et al. \cite{xiaoevolutive} & ES\\
Zhao et al. \cite{zhao2022elite} & policy learning \\
Fu et al. \cite{fu2023communication} & hierarchical multi-agent RL \\
Al-Hamid and Al-Anbuky \cite{al2022vehicular} & N/A\\
Jiao et al. \cite{jiao2023mobile} & N/A\\
Ren et al. \cite{ren2023end} & GraphSAGE, DDQN, BCNN \\
Rosello et al. \cite{rosello2023network} & N/A\\
Kherbache et al. \cite{kherbache2022digital} & N/A\\
Wei et al. \cite{wei2021data} & N/A\\
Raj et al. \cite{raj2023building} & KG, Template-based contextual information \\
\botrule
\end{tabular*}}{}
\end{table}

\subsubsection{Resource Allocation}
In the realm of network management through efficient resource allocation, various cutting-edge approaches have emerged. For cellular networks, Sun et al. \cite{sun2021digital} initiate this narrative by introducing a DTN designed for 5G technology. Their work focuses on upholding stringent SLA standards, a crucial aspect of network performance. This approach ultimately boosts network efficiency by facilitating comprehensive mapping and management. Baranda et al. \cite{baranda2021aiml} introduce an AI/ML platform integrated into the 5G workflow for scaling based on real-time data metrics, catering to SLA management in a DT service scenario. Zhou et al. \cite{zhou2022digital} introduce a resource management scheme that uses DT technology for 6G. In \cite{huang2023collective}, Huang et al. focus on developing a collective RL method to efficiently allocate resources in real-time and adapt to varying service demands. Tao et al. \cite{tao2023drl} delve into resource management within DTNs for 6G service requests. They focus on enhancing service response using a software-defined DTN architecture and Proximal Policy Optimization DRL (PPO-DRL). Duran et al. \cite{duran2023digital} aim to strengthen core network management efficiency by integrating intelligent methods into topology discovery processes. The technique reduces complexity and resource consumption, mainly employing Multilayer Perceptron (MLP) for visit decision recommendations. Vila et al. \cite{vila2023design} introduce an NDT architecture tailored for RANs within 5G networks. Their approach leverages RL to train and optimize RAN operations through the NDT framework. 

Su et al. \cite{su2022digital} shift the focus toward addressing compute-intensive applications within MEC environments. They introduce a DT-based task offloading scheme, leveraging the potent Double DQN (DDQN) approach. This novel strategy optimizes resource allocation, enhancing network performance. Also, for MEC, Dai et al. \cite{dai2022service} bring forward a DTN-assisted system designed to optimize service placement and workload distribution. Their primary focus revolves around the strategic placement of services and efficient distribution of workloads within these systems. He et al. \cite{he2022resource} introduce a hierarchical FL framework, seamlessly integrating DT and MEC into cellular networks. This approach optimizes resource allocation and network performance, featuring DRL. Guo et al. \cite{guo2022time} introduce a network sensing edge deployment optimization mechanism for DT systems. The mechanism enhances management efficiency by optimizing edge deployment based on network state, employing an activity estimation model and a chaotic Particle Swarm Optimization (PSO) algorithm. Merging DT and MEC concepts, Yuan et al. \cite{yuan2023joint} use DNN for task offloading with A3C, aiming to optimize traffic in pursuit of reduced inference latency. Luo et al. \cite{luo2022distributed} harness DT technology to boost efficiency in wireless communication networks. Their central focus lies in optimizing resource allocation through distributed DRL techniques. Wieme et al. \cite{wieme2022relay} optimize Bluetooth mesh networks with DT technology, emphasizing efficient relay selection. Their work underscores the pivotal role of AI-driven optimization techniques in enhancing network behavior and configuration.

For SDNs, Naeem et al. \cite{naeem2021digital} propose a DT-enabled Deep Distributional Q-network (DDQN) framework to optimize resource allocation and network slicing policies. Abdel-Basset et al. \cite{abdel2023digital} delve into slicing-enabled communication networks, aiming for optimal resource allocation while maintaining QoS. Their solution relies on FL and differential privacy techniques. Likewise, Hong et al. \cite{hong2021netgraph} present the NetGraph DT platform for intelligent data center network management, and in \cite{lombardo2022design} a platform combining SDN and DT technology for autonomous network management is introduced. Similarly, Gong et al. \cite{gong2022resource} propose a holistic network virtualization architecture that combines DT and network slicing. They introduce an environment-aware offloading mechanism based on integrated sensing and communication systems to address computation offloading challenges. The use of AI, including the Shapley-Q value and DDPG algorithm, aids in solving optimization problems related to task scheduling and resource allocation.

In IIoT and IoT networks, Lu et al. \cite{lu2020communication} merge DTs with edge networks, creating DTNs for IoT optimization. They employ blockchain-empowered FL and RL for traffic pattern optimization and resource allocation. Dai et al. \cite{dai2020deep} employ DTN and the Lyapunov optimization method with the A3C algorithm to enhance energy efficiency and processing in IIoT systems by optimizing resource allocation. Bellavista et al. \cite{bellavista2021application} propose an Application-driven DTN middleware for IIoT. Their focus is on simplifying device interactions and dynamically managing network resources. Luan et al. \cite{luan2022optimization} develop an intelligent industrial system that synergizes AI and DTs, ultimately optimizing network performance in smart sensors for manufacturing. AI-driven load-balancing strategies are among their tools to improve network performance. Guo et al. \cite{guo2022federated} propose a device-to-device communication-aided DT-edge network for efficient management in 6G IIoT. Tang et al. \cite{tang2023digital} delve into the realm of Industry 4.0, where they utilize DTN technology to personalize services. AI techniques like Multi-Agent Deep Deterministic Policy Gradients (MA-DDPG) are instrumental in optimizing resource allocation for personalized IIoT services. Geisler et al. \cite{geissler2023mvnocoresim} manage overload in IoT mobile networks, specifically focusing on the intricate task of orchestrating signaling traffic.

Morette et al. \cite{morette2023machine} optimize optical networks by harnessing the power of DT technology, emphasizing the pivotal role of ML techniques in elevating network performance and the strategic allocation of vital resources. In parallel, \cite{curri2023digital, borraccini2023experimental} utilize a DT-based approach at the physical layer of optical networks to evaluate transmission quality and overall network performance. Hao et al. \cite{hao2023intelligent} focus on automating the scheduling and maintenance of optical transmission networks. The DTN-based automatic scheduling method uses a modified DDPG algorithm for topology optimization, aiming to reduce maintenance costs and improve resource utilization. Wu et al. \cite{wu2023dynamic} introduce a multifactor-associated Network Topology Portrait (NTP) scheme for DTN in optical networks. The scheme aims to optimize dynamic routing computation using different routing algorithms, emphasizing improving efficiency and performance.

Fu et al. \cite{fu2022alsodtn} aim to enhance air logistics through DTN with AI-driven optimization for satellite and aerial-based networks. The framework incorporates transformer-based information fusion and multi-agent DRL for UAV cooperation in route planning. In emergency communication scenarios, T. Guo \cite{guo2022connectivity} focuses on establishing seamless connections between users and resource-constrained aerial base stations using device-to-device communication and advanced Q-learning techniques. Gong et al. \cite{gong2023computation} pioneer innovative approaches, employing a blockchain-aided Stackelberg game model and a Lyapunov stability theory-based model-agnostic meta-learning framework. These methodologies are strategically applied to foster optimal resource allocation and informed decision-making within the intricate landscape of satellite-ground integrated DTNs. Zhang et al. \cite{zhang2023uav} fine-tune dynamic data transmission within the expansive domain of DT services. UAVs take center stage as AI techniques deftly chart optimal flight paths and refine data transmission strategies, effectively elevating network performance and enhancing user satisfaction. Likewise, Gong et al. \cite{gong2023computation} venture into the dynamic realm of UAV-assisted edge computing systems. Their work, rooted in DRL, is dedicated to the meticulous optimization of service migration processes, promising to augment network efficiency further. 

For vehicular networks, Dai and Zhang \cite{dai2022adaptive} propose enhancing vehicular edge computing networks using adaptive DT-enabled networks and employing DRL to minimize offloading latency. Li et al. \cite{li2022digital} introduce a novel approach for computing resource management of edge servers in vehicular networks. They emphasize constructing tailored two-tier DTs and employing AI for real-time resource allocation. Cazzella et al. \cite{cazzella2023multi} contribute a data-driven approach to create a DT for V2X communication scenarios emphasizing addressing high mobility challenges. These diverse approaches illuminate the path toward efficient resource allocation in the ever-evolving network management landscape. Researchers have collectively paved the way for innovative solutions, leveraging DT technology and AI techniques to optimize resource allocation from cellular networks to IoT, optical networks, air logistics, and vehicular networks.

\begin{table}[!tbhp]
\processtable{Resource Allocation Works}
{\begin{tabular*}{\textwidth}{@{\extracolsep{\fill}}ll@{}}\toprule
\textbf{Work} & \textbf{Used AI Tools} \\
\midrule
Sun et al. \cite{sun2021digital} & DTN, SLA mapping \\
Baranda et al. \cite{baranda2021aiml} & ML\\
Zhou et al. \cite{zhou2022digital} & N/A\\
Huang et al. \cite{huang2023collective} & Collective RL \\
Tao et al. \cite{tao2023drl} &PPO-DRL \\
Duran et al. \cite{duran2023digital} & MLP \\
Vila et al. \cite{vila2023design} & RL \\
Su et al. \cite{su2022digital} &DDQN \\
Dai et al. \cite{dai2022service} & N/A\\
He et al. \cite{he2022resource} & Hierarchical FL, DRL \\
Guo et al. \cite{guo2022time} & N/A\\
Yuan et al. \cite{yuan2023joint} & DNN, A3C \\
Luo et al. \cite{luo2022distributed} & Distributed DRL \\
Wieme et al. \cite{wieme2022relay} & N/A\\
Naeem et al. \cite{naeem2021digital} & DDQN \\
Abdel-Basset et al. \cite{abdel2023digital} & FL, differential privacy \\
Hong et al. \cite{hong2021netgraph} & NetGraph\\
Gong et al. \cite{gong2022resource} & N/A\\
Lu et al. \cite{lu2020communication} & blockchain-empowered FL, RL \\
Dai et al. \cite{dai2020deep} & DTN, Lyapunov optimization, A3C algorithm \\
Bellavista et al. \cite{bellavista2021application} & middleware \\
Luan et al. \cite{luan2022optimization} & N/A\\
Guo et al. \cite{guo2022federated} & N/A\\
Tang et al. \cite{tang2023digital} & MADDPG \\
Geisler et al. \cite{geissler2023mvnocoresim} & N/A\\
Morette et al. \cite{morette2023machine} & ML techniques \\
Curri et al. \cite{curri2023digital}, Borraccini et al. \cite{borraccini2023experimental} & N/A\\
Hao et al. \cite{hao2023intelligent} & automatic scheduling, DDPG \\
Wu et al. \cite{wu2023dynamic} & Multifactor-associated NTP \\
Fu et al. \cite{fu2022alsodtn} & transformer-based fusion, multi-agent DRL\\
T. Guo \cite{guo2022connectivity} & Q-learning \\
Gong et al. \cite{gong2023computation} & blockchain, Stackelberg model, Lyapunov stability,\\
&meta-learning\\
Zhang et al. \cite{zhang2023uav} & N/A \\
Liu et al. \cite{liu2021digital} & Decision Tree, double DL \\
Dai and Zhang \cite{dai2022adaptive} & DRL \\
Li et al. \cite{li2022digital} & N/A\\
Cazzella et al. \cite{cazzella2023multi} & N/A\\
\botrule
\end{tabular*}}{}
\end{table}

\subsection{AI-Based Communication Enhancement}\label{communicationenhancement}
Researchers have made several notable contributions to communication enhancement through DTN by employing AI, ML, and DL techniques. Sun et al. \cite{sun2020dynamic} introduce the DT edge network concept for mobile offloading decision-making in 6G environments; their contribution utilizes Lyapunov optimization and the A3C DRL algorithm. Wang et al. \cite{wangdigital} develop a DTN for UAV swarm-based 5G emergency networks. They leverage DL techniques for swarm deployment under varying conditions to enhance communication efficiency during emergencies. Jian et al. \cite{jian2021study} propose DT technology for communication channels beyond 5G and 6G communication systems. Xiang et al. \cite{xiang20225g} create a 5G wireless network DT system using ray tracing propagation modeling. The system accurately models wireless signals for 5G networks, helping identify weak coverage areas and optimize engineering parameters. In the IIoT context, \cite{lu2020communication,zhao2022communication} introduce DT edge networks of FL to optimize communication efficiency and reduce transmission energy costs. Liang et al. \cite{liang2022digital} address the limitations of traditional wavelength division multiplexing communication networks for power IoT with an electric-elastic optical networks architecture. Their work optimizes communication bandwidth scheduling and introduces a UCB automatic routing selection algorithm. Li et al. \cite{li2023adaptive} study a DT-empowered integrated sensing, communication, and computation network, optimizing Multi-Input Multi-Output (MIMO) radars and computation offloading energy consumption. They employ the Multiagent Proximal Policy Optimization (MAPPO) framework for intelligent offloading decisions.

Regarding V2X communications, Zelenbaba et al. \cite{zelenbaba2022wireless} use DTs to evaluate hardware and system performance in wireless vehicular communication links. They create site-specific DTs to assess link reliability in vehicular communication scenarios. Lv et al. \cite{lv2022deep} optimize optical wireless communications in Intelligent Transportation Systems (ITS) with a focus on Visible Light Communications (VLC). They propose a carrier-less amplitude/phase modulation scheme and employ Convolutional Neural Networks (CNNs) for feature extraction. Demir et al. \cite{demirdigital} apply DTs in connected and autonomous vehicles to improve wireless QoS in non-line-of-sight scenarios. Their methodology predicts and mitigates the impact of obstacles on wireless communication. Liu et al. \cite{liu20236g} focus on achieving energy-efficient communication in the IoV within 6G mobile networks. They introduce a DT method combined with ML to model the millimeter-wave channel and optimize energy-efficient communication. These contributions underscore the significant role of AI, ML, and DL techniques within DTNs in enhancing communication efficiency and addressing emerging challenges in various communication scenarios.

\begin{table}[!tbhp]
\processtable{Communication Enhancement Works}
{\begin{tabular*}{\textwidth}{@{\extracolsep{\fill}}ll@{}}\toprule
\textbf{Work} & \textbf{Used AI Tools} \\
\midrule
Sun et al. \cite{sun2020dynamic} & Lyapunov optimization, A3C DRL \\
Wang et al. \cite{wangdigital} & DL \\
Jian et al. \cite{jian2021study} &N/A \\
Xiang et al. \cite{xiang20225g} & Ray tracing propagation modeling \\
\cite{lu2020communication,zhao2022communication} & FL \\
Liang et al. \cite{liang2022digital} &  UCB automatic routing selection algorithm \\
Li et al. \cite{li2023adaptive} & MAPPO\\
Zelenbaba et al. \cite{zelenbaba2022wireless} & N/A \\
Lv et al. \cite{lv2022deep} & CNN \\
Demir et al. \cite{demirdigital} & N/A \\
Liu et al. \cite{liu20236g} & ML \\
\botrule
\end{tabular*}}{}
\end{table}

\subsection{AI-Based Prediction Analysis}\label{prediction}
DTNs have gained prominence recently for their ability to address various prediction analysis tasks across diverse domains. These tasks often involve using AI, ML, and DL techniques to enhance predictions and optimize network performance. A collection of works that employ DTNs to tackle prediction challenges has been presented. To predict Key Performance Indicators (KPIs), Wang et al. \cite{wang2020graph} utilize DT technology to manage the complexities of slicing in 5G networks. They create virtual representations of slicing-enabled networks and predict performance changes. They hint at using the GNN model to capture relationships among slices and predict metrics like end-to-end latency. Schippers et al. \cite{schippers2023data} introduce a novel DT approach to predict KPIs for mission-critical vehicular applications. They employ AI techniques to accurately predict KPIs such as data rate and latency, enhancing QoS predictions for smart city services.
Similarly, Baert et al. \cite{baert2021digital} leverage DT technology to optimize Bluetooth mesh networks for IoT applications, using AI-driven DT to predict and optimize end-to-end latencies, packet delivery ratios, and path distributions. Ferriol-Galmes et al. \cite{ferriol2022flowdt} enhance network modeling using GNNs and DL techniques, modeling computer networks at the fine-grained flow level to predict per-flow KPIs accurately. Saravanan et al. \cite{saravanan2022enabling} build a scalable NDT to predict per-path mean delay in large-scale communication networks, exploring various Neural Network (NN) architectures, including GNNs, to enhance prediction accuracy. Padmapriya and Srivenkatesh \cite{padmapriya2023digital} focus on improving the functionality and maintenance of smart home gadgets using the Deep CNN Logistic Regression Model with DTs to predict gadget functionality performance. Fu et al. \cite{fu2022digital} propose a time delay prediction algorithm for vehicular networks using NNs, aiming to address performance degradation caused by network latency. Li et al. \cite{li2023learnable} propose a learning-based NDT for efficiently estimating wireless network configuration KPIs before physical implementation. He et al. \cite{he2023conditional} optimize wireless networks with massive MIMO technology using a DTN approach, employing a Conditional GAN (C-GAN) for accurate KPI predictions and pre-validation.

For traffic prediction, Lai et al. \cite{lai2023deep} develop a DL approach in DTNs, introducing the eConvLSTM model for predicting background traffic matrices in LANs, improving prediction accuracy significantly. Morette et al. \cite{morette2023machine} enhance traffic QoT estimation and prediction in optical NDTs using ML, introducing an NN architecture for accurate predictions across different configurations and devices. Nie et al. \cite{nie2023digital} develop a network traffic prediction algorithm for vehicular networks using Deep Q-learning (DQN) and Generative Adversarial Networks (GAN). In terms of ITS traffic prediction, Ji et al. \cite{ji2022digital} predict the spatiotemporal congestion resulting from traffic accidents in urban road networks using a Convolutional LSTM (Conv-LSTM) network, focusing on macroscopic traffic operation. Xu et al. \cite{xu2023traffnet} introduce TraffNet, a DL framework for road NDTs that considers the causality of traffic volume from vehicle trajectory data to improve traffic prediction accuracy for just-in-time decision-making. Likewise, Dangana et al. \cite{dangana2022towards} develop a prototype system for metamorphic object transportation, focusing on predicting the transportation process of deformable interlinked linear objects.

These studies collectively demonstrate the versatility of DTNs in addressing prediction analysis tasks across different domains and the role of AI, ML, and DL techniques in improving prediction accuracy and network optimization.

\begin{table}[!tbhp]
\processtable{Prediction Analysis Works}
{\begin{tabular*}{\textwidth}{@{\extracolsep{\fill}}ll@{}}\toprule
\textbf{Work} & \textbf{Used AI Techniques} \\
\midrule
Wang et al. \cite{wang2020graph} & GNN \\
Schippers et al. \cite{schippers2023data} & N/A \\
Baert et al. \cite{baert2021digital} & N/A \\
Ferriol-Galmes et al. \cite{ferriol2022flowdt} & GNNs, DL \\
Saravanan et al. \cite{saravanan2022enabling} & NN, GNNs \\
Padmapriya and Srivenkatesh \cite{padmapriya2023digital} & Deep CNN Logistic Regression Model \\
Fu et al. \cite{fu2022digital} & NN\\
Li et al. \cite{li2023learnable} & N/A \\
He et al. \cite{he2023conditional} & C-GAN \\
Lai et al. \cite{lai2023deep} & eConvLSTM model \\
Morette et al. \cite{morette2023machine} & ML, NN architecture \\
Nie et al. \cite{nie2023digital} & DQN, GAN \\
Ji et al. \cite{ji2022digital} & Conv-LSTM \\
Xu et al. \cite{xu2023traffnet} & DL\\
Dangana et al. \cite{dangana2022towards} & N/A\\
\botrule
\end{tabular*}}{}
\end{table}

\subsection{AI-Based Fault and Anomaly Detection}\label{anomaly}
In fault and anomaly detection, recent academic literature has witnessed a proliferation of research endeavors that leverage DTNs. Towards fault detection, Zhu et al. \cite{zhu2023fault} focus on 5G networks where the scarcity of fault samples and monitoring data has been a persistent hurdle. They use the Average Wasserstein GAN with Gradient Penalty (AWGAN-GP) to enhance failure discovery and prediction. The XGBoost algorithm is harnessed for real-time fault localization within the physical network. Zheng et al. \cite{zheng2022practice} introduce a fault self-recovery method targeted at 5G networks. Their methodology employs a data governance approach to construct models representing physical network components in a DT. Visual topology technology is deployed to extract Knowledge-as-a-Service (KaaS) capabilities that facilitate call quality tests, fault-propagation chain reasoning, and disaster recovery analysis. Mayer et al. \cite{mayer2022demonstration} demonstrate an ML-based framework for localizing soft failures in optical transport networks. Their focus is on harnessing SDN and streaming-based telemetry to automate detecting soft shortcomings that may not trigger conventional alarms. Their experimental setup entails an SDN-controlled network with transponders and an optical line system providing telemetry data. Artificial NN ML plays a pivotal role in their failure localization. Similarly, Wang and Chen \cite{wang2022design} optimize the backbone optical transport network. Their work introduces the concept of DTs to address scenarios like equipment defects, fault warnings, and trend analysis.

On the other hand, Calvo-Bascones et al. \cite {calvo2023collaborative} propose an anomaly detection methodology tailored specifically for industrial systems with the concept of a Snitch DT, which models the connections between physical behaviors to detect anomalies. Leveraging spatiotemporal features and quantile regression, they characterize the behavior of individual physical entities. Li et al. \cite{li2022anomaly} aim to combat Internet service quality degradation by proposing a DT system capable of quasi-real-time simulation for comprehensive indicator analysis, anomaly detection, and intelligent operation. In a slightly different vein, Zhu et al. \cite{zhu2021impact} pivot their research toward analyzing the causes and effects of noise and distortion in the DT context, particularly focusing on AI-driven optical networks. Liu et al. \cite{liu2021distributed} contribute to amalgamating DT technology into vehicular edge networks to support cybersecurity and anomaly detection within vehicular networks. Their primary thrust revolves around establishing a distributed trust evaluation system. Kaytaz et al. \cite{kaytaz2023graph} put forth a pioneering GNN anomaly detection framework tailored to ensuring reliability within multi-dimensional data streams within ITS. Their approach employs unsupervised and supervised ML techniques, including BiDirectional Generative Adversarial Networks (BiGAN), affinity propagation, and Graph CNN, to model data streams and perform anomaly detection.

\begin{table}[!tbhp]
\processtable{Fault and Anomaly Detection Works}
{\begin{tabular*}{\textwidth}{@{\extracolsep{\fill}}ll@{}}\toprule
\textbf{Work} & \textbf{Used AI Techniques} \\
\midrule
Zhu et al. \cite{zhu2023fault} & AWGAN-GP, XGBoost \\
Zheng et al. \cite{zheng2022practice} & Data governance, Visual topology \\
Mayer et al. \cite{mayer2022demonstration} & ML, ANN\\
Wang and Chen \cite{wang2022design} & N/A \\
Calvo-Bascones et al. \cite{calvo2023collaborative} & Snitch DT, spatiotemporal features, quantile regression \\
Li et al. \cite{li2022anomaly} & N/A \\
Zhu et al. \cite{zhu2021impact} & N/A \\
Liu et al. \cite{liu2021distributed} & Distributed Trust Evaluation \\
Kaytaz et al. \cite{kaytaz2023graph} & GNN, BiGAN, affinity propagation, Graph CNN \\
\botrule
\end{tabular*}}{}
\end{table}

\subsection{AI-Based Security and Privacy Preservation}\label{security}
DTNs have witnessed significant growth and innovation in addressing security and privacy challenges in recent years. This literature review explores the common themes and key contributions across various works that utilize AI, ML, and DL techniques to enhance security and privacy in the context of DTNs.

Dong et al. \cite{dong2021dual} introduce a bidirectional mapping between physical and virtual spaces within DT technology. Their dual blockchain framework enhances data security in DT scenarios, emphasizing the importance of trustworthiness. Kumar and Khari \cite{kumar2021architecture} propose a DT-based framework for network forensic analysis to detect and prevent cyberattacks. The integration of tools like Nmap and Wireshark suggests AI utilization. Son et al. \cite{son2022design} focus on secure sharing of DT data using cloud computing and blockchain, emphasizing privacy preservation and data security in wireless channels. They employ formal methods like BAN logic and the AVISPA simulation tool for security analysis. Qu et al. \cite{qu2022fedtwin} address DTN challenges through a blockchain-enabled adaptive asynchronous FL paradigm. Their work integrates AI techniques, blockchain, and consensus algorithms to enhance privacy, security, and reliability in DTNs. Zhan et al. \cite{zhan2022implementation} tackle practical DT deployment in cloud-native networks. Su and Qu \cite{su2022detection} introduce an intrusion detection method using FL and LSTM models for network traffic analysis within DTNs. Their approach enhances privacy and accuracy and outperforms existing methods in real-time monitoring. Yigit et al. \cite{yigit2022digital} propose an intelligent DDoS detection mechanism for core networks. R. Bagrodia \cite{bagrodia2023using} emphasizes assessing cyber resiliency in defense systems, particularly in complex scenarios. They build a DTN to analyze network operations and vulnerabilities to counter potential threats.

In the rapidly evolving landscape of cellular networks, Lu et al. \cite{lu2020low} integrate DTNs and 6G networks to fortify wireless connectivity in IIoT. They use FL and blockchain to enhance reliability, security, and data privacy. They also leverage multiagent RL for optimized edge association, showcasing tangible efficiency improvements and cost reductions. Vakaruk et al. \cite{vakaruk2021digital} use DTNs and ML tools to train cybersecurity experts. Chen et al. \cite{chen2021convoy_dtn} introduce the DT protection function, a security interaction engine that safeguards message confidentiality, integrity, stability, and non-reputation while disseminating policies within DTNs. Wang et al. \cite{wang2022digital} propose a DT-based approach to autonomously provision security functions for 5G network slices. The core objective is dynamically allocating security capabilities based on KPIs, ensuring real-time network slice security. Ozdogan et al. \cite{ozdogan2022digital} leverage ML techniques to optimize 6G network parameters and address privacy concerns. They integrate Blockchain and Transfer Learning (TL) to maintain data security and privacy during network recovery and expansion.

Addressing security and privacy tasks in the context of DTNs within IoT and IIoT environments, Kumar et al. \cite{kumar2022blockchain} integrate blockchain and DL techniques like LSTM Sparse AutoEncoder and Multi-Head Self-Attention (MHSA)-based Bidirectional Gated Recurrent Unit (BiGRU) algorithms to enhance communication security, detect attacks, and improve data privacy. Lv et al. \cite{lv2022digital} integrate quantum communication techniques with DTs to enhance IIoT communication security. They propose a channel encryption scheme that uses quantum communication principles to ensure secure IIoT communication. Grasselli et al. \cite{grasselli2022industrial} enable experimentation in DTNs with network topologies, attacks, and countermeasures without impacting real cyber-physical systems. Feng et al. \cite{feng2023game} adopt a game theory approach to enhance the security of industrial DTNs. Their study explores vulnerability mining and repair using evolutionary game theory, cooperative strategies for vulnerability patch development, and the integration of differential evolution into the Wolf Colony Algorithm (WCA). Danilczyk et al. \cite{danilczyk2021blockchain} propose a novel implementation of the SHA-256 hash algorithm for creating a blockchain of sensor readings in a DT environment, focusing on secure two-party communications. Their work ensures data integrity and security through a chained checksum approach. Feng et al. \cite{feng2021sensible} involve interference source location schemes, encryption techniques, and filtering methods to reduce interference attacks in IoT-based DT communication networks. Jiang et al. \cite{jiang2021cooperative} propose a blockchain-based DT edge networks framework for secure and flexible DT construction in IoT. Their approach utilizes cooperative FL and blockchain, emphasizing secure model updates and efficient resource allocation. Bozkaya et al. \cite{bozkaya2023proof} also introduce a holistic approach integrating DT and blockchain into edge networks to optimize task scheduling in IoT applications. Their Proof of Evaluation (PoE)-based algorithm employs genetic algorithms to balance energy efficiency and delay reduction while ensuring data integrity. Qian et al. \cite{qian2023secrecy} focus on constructing a secure DT for the Marine IoT using FL and non-orthogonal multiple access. Kherbache et al. \cite{kherbache2023iot} address the security and resilience of information and weapon systems against cyber threats, particularly in complex scenarios. Lv and Qiao \cite{lv2022context} explore sustainable DTs and context-aware computing to enhance network security in industrial environments. Wang et al. \cite{wang2023digital} offer a comprehensive exploration of security challenges within DT-enabled wireless systems. Their framework utilizes AI techniques like game theory and ML to enhance the system's security against attacks and flawed parameters.

Wu et al. \cite{wu2022deep} explore security concerns and computational intelligence in a drone information system employing DL techniques to fortify the defenses in the evolving landscape of UAV-based networks. Their work focuses on predictive modeling, leveraging an improved LSTM network to analyze data, specifically control signal data, to predict potential attacks on the drone system. They introduce differential privacy frequent subgraphs and utilize DTs technology to map the drone's physical operating environment to maintain data privacy. Likewise, Dai et al. \cite{dai2022digital} leverage DTs to elevate strategy optimization and decision-making within air-ground integrated networks, particularly in urban sensing and disaster relief applications. The study explores an aerial blockchain-based approach to fortify data security within secure federated aerial learning. He et al. \cite{he2023federated} introduce a Federated Continuous Learning framework with a Stacked Broad Learning System (FCL-SBLS) tailored for intrusion detection systems in edge UAV-based IoT. They aim to enhance security and privacy while training intrusion detection models with distributed UAV data. The proposed framework, assisted by the DTN, facilitates continuous learning and training of models through asynchronous FL and a DDPG-based UAV selection scheme. AI components like federated learning and DDPG for UAV selection are pivotal in optimizing the IDS model's performance.

In V2X-based networks, several scholarly endeavors tackle the pressing security and privacy issues. Liu et al. \cite{liu2021distributed} enhance cybersecurity within vehicular networks by establishing a distributed trust evaluation system aided by DTs. Xu et al. \cite{xu2021efficient} introduce a cloud-based DT that synchronizes with the corresponding Autonomous Vehicle (AV) in real-time, fostering information fusion and computing. Moreover, their work presents a concrete authentication protocol \cite{al2019data} to ensure secure communication between AVs and DTs, underscoring the significance of communication security and privacy protection. Similarly, Liu et al. \cite{liu2022blockchain} propose blockchain technology and DTs to fortify IoV security. Their work encompasses creating a secure communication architecture for IoV based on immutable and traceable blockchain data.
Moreover, they develop a risk forecast model for IoV node security utilizing Wasserstein Distance GAN (WaGAN) models, showcasing the accelerated learning rates due to Wasserstein distance. Zhang et al. \cite{zhang2022bsdp} propose a blockchain-based smart parking scheme within DT-empowered networks to monitor and predict traffic conditions around parking lots. This scheme leverages blockchain and smart contracts to ensure reliable data storage and correct parking responses while prioritizing privacy protection for drivers. He et al. \cite{he2022security} delve into the intricacies of vehicular DTNs within the context of AVs, highlighting security and privacy concerns. Their work primarily focuses on the challenges posed by the collection of sensitive user information and exposure to open network environments. Yang et al. \cite{yang2022privacy} introduce a combined multi-armed bandit-based auction incentive mechanism designed to identify the quality of participants in the vehicular edge network without compromising sensitive information. This mechanism aims to enhance the efficiency and accuracy of the DT model within vehicular edge networks. Collectively, these works underscore the transformative potential of integrating DTNs into V2X networks, with AI, ML, and DL techniques serving as integral components for strengthening security, privacy, and network efficiency.

\begin{table}[!tbhp]
\processtable{Security and Privacy Works with AI Tools}
{\begin{tabular*}{\textwidth}
{@{\extracolsep{\fill}}ll@{}}\toprule
\textbf{Work} & \textbf{Used AI Tools} \\
\midrule
Dong et al. \cite{dong2021dual} & Dual blockchain \\
Kumar and Khari \cite{kumar2021architecture} & Nmap, Wireshark\\
Son et al. \cite{son2022design} & BAN logic, AVISPA, AI, blockchain \\
Qu et al. \cite{qu2022fedtwin} & Blockchain, FL, consensus algorithms \\
Zhan et al. \cite{zhan2022implementation} & N/A \\
Su and Qu \cite{su2022detection} & FL, LSTM \\
Yigit et al. \cite{yigit2022digital} & N/A\\
R. Bagrodia \cite{bagrodia2023using} & N/A \\
Lu et al. \cite{lu2020low} & FL, blockchain, multiagent RL \\
Vakaruk et al. \cite{vakaruk2021digital} & ML \\
Chen et al. \cite{chen2021convoy_dtn} &N/A\\
Wang et al. \cite{wang2022digital} & N/A\\
Ozdogan et al. \cite{ozdogan2022digital} & ML, blockchain, TL \\
Kumar et al. \cite{kumar2022blockchain} & Blockchain, DL  \\
Lv et al. \cite{lv2022digital} & Quantum  \\
Kumar et al. \cite{kumar2022blockchain} & Blockchain, LSTM Sparse AutoEncoder, MHSA-based BiGRU \\
Lv et al. \cite{lv2022digital} & Quantum, encryption \\
Grasselli et al. \cite{grasselli2022industrial} & N/A \\
Feng et al. \cite{feng2023game} & Game theory, evolutionary game theory, differential evolution \\
Danilczyk et al. \cite{danilczyk2021blockchain} & SHA-256, blockchain, chained checksum \\
Feng et al. \cite{feng2021sensible} & Interference source location schemes, encryption \\
Jiang et al. \cite{jiang2021cooperative} & Blockchain, cooperative FL \\
Bozkaya et al. \cite{bozkaya2023proof} &blockchain, genetic algorithm\\
Qian et al. \cite{qian2023secrecy} & FL, non-orthogonal multiple access \\
Kherbache et al. \cite{kherbache2023iot} & N/A \\
Lv and Qiao \cite{lv2022context} & N/A\\
Wang et al. \cite{wang2023digital} & Game theory, ML \\
Wu et al. \cite{wu2022deep} & DL, LSTM, differential privacy frequent subgraphs \\
Dai et al. \cite{dai2022digital} & blockchain \\
He et al. \cite{he2023federated} & FCL-SBLS, asynchronous FL, DDPG\\
Liu et al. \cite{liu2021distributed} & Distributed trust\\
Xu et al. \cite{xu2021efficient} & authentication \\
Liu et al. \cite{liu2022blockchain} & Blockchain, WaGAN model \\
Zhang et al. \cite{zhang2022bsdp} & Blockchain, smart contracts \\
He et al. \cite{he2022security} & N/A\\
Yang et al. \cite{yang2022privacy} & Multi-armed bandit-based auction incentive mechanism \\
\bottomrule
\end{tabular*}}{}
\end{table}

\section{Main AI Models and Tools Harnessed by DTNs}
In our analysis of the literature reviewed in the previous section, we observed the extensive utilization of AI models, techniques, and tools to address the mentioned tasks within DTNs. Ordered by their frequency of use in contemporary academic research, the following list offers valuable insights into the pivotal technologies that are shaping the future of DTNs. They are mainly ML, DL, RL, FL, and Graph-based.
\subsection{Machine Learning (ML) Tools and Models}
In DTNs, ML tools enable predictive analysis, enhance security and privacy, optimize task offloading, and drive energy efficiency. These tools encompass a range of techniques, from traditional methods to more advanced algorithms, each tailored to address specific challenges within the digital twin framework \cite{mayer2022demonstration, bozkaya2023proof, zhao2022interlink, liu2023deep, mayer2022demonstration, ozdogan2022digital, vakaruk2021digital, wang2023digital, ursu2023towards, ozdogan2022digital, zhu2023fault, fu2022digital, saravanan2022enabling}

Neural Networks (NNs) are a fundamental ML tool widely applied in DTNs. Mayer et al. \cite{mayer2022demonstration}, Fu et al. \cite{fu2022digital}, and Saravanan et al. \cite{saravanan2022enabling} have employed NNs for prediction analysis. These versatile models demonstrate their strength in understanding complex data patterns and temporal dependencies, making them indispensable for accurate predictions.
\begin{boxes}
{\boxhead{Neural Networks (NN)}}
{\textit{Neural Networks are models inspired by the human brain's structure and functioning. They consist of interconnected nodes organized in layers to process and learn from data.}}
\end{boxes}

Genetic Algorithms, as showcased by Bozkaya et al. \cite{bozkaya2023proof} and Zhao et al. \cite{zhao2022interlink}, step into the spotlight with a focus on security, privacy, and task optimization. Genetic algorithms are trained to explore search spaces to find optimal solutions. Bozkaya et al. utilize genetic algorithms to strengthen security and privacy measures within DTNs. Zhao et al., on the other hand, leverage genetic algorithms for task offloading and content delivery optimization, optimizing resource allocation and enhancing content delivery efficiency.

\begin{boxes}
{\boxhead{Genetic Algorithms}}
{\textit{Genetic Algorithms are optimization algorithms inspired by natural selection and genetics. They are used to find solutions to optimization and search problems. Genetic Algorithms maintain a population of potential solutions and evolve them over generations through selection, crossover (recombination), and mutation.}}
\end{boxes}

Transfer Learning (TL) finds its application in enhancing security and privacy within DTs \cite{ozdogan2022digital}. TL allows for knowledge transfer from one domain to another, improving the robustness of security measures and safeguarding sensitive information.

\begin{boxes}{\boxhead{Transfer Learning (TL)}} {\textit{TL is an ML technique in which a pre-trained model, typically trained on a large dataset, is adapted for a different but related task. Instead of training a model from scratch, TL leverages the knowledge and feature representations learned by the pre-trained model. This approach can significantly reduce training time and data requirements for the target task and is especially useful when labeled data is scarce or expensive to obtain.}} \end{boxes}

Finally, the Extreme Gradient Boost (XGBoost) algorithm is harnessed by Zhu et al. \cite{zhu2023fault} for prediction analysis. XGBoost excels at handling structured data and enhancing prediction accuracy, making it an ideal choice for data-driven predictions in DT systems.

\begin{boxes}{\boxhead{Extreme Gradient Boost (XGBoost)}} {\textit{XGBoost is a popular ML algorithm that belongs to the ensemble learning category. It is designed to handle various supervised learning tasks, including classification and regression. XGBoost combines the predictions of multiple weak learners into a strong, robust model. It uses a gradient boosting framework, which optimizes model performance by iteratively adding new trees that correct the errors made by previous ones.}} \end{boxes}

\subsection{Deep Learning (DL) Models and Techniques}
DL has emerged as an indispensable tool for tackling multifaceted challenges within DTNs. In our comprehensive literature review, DL is pivotal in enhancing network performance and managing various network aspects, encompassing planning, resource allocation, communication refinement, traffic and KPIs prediction, anomaly detection, and reinforcing network security and privacy \cite{zhu2023fault,kaytaz2023graph,ji2022digital,padmapriya2023digital,ferriol2022flowdt,kumar2022blockchain,wu2022deep,han2022polymorphic,kaytaz2023graph,su2022detection,wu2022deep,yi2022digital,liu2022blockchain}.

Convolutional Neural Networks (CNNs) stand at the forefront of this technological advancement. These specialized models excel at processing grid-like data, particularly images. Their application in DTs, as demonstrated by Padmapriya and Srivenkatesh \cite{padmapriya2023digital}, revolves around prediction analysis. By leveraging the inherent capabilities of CNNs, researchers have successfully harnessed image-based data to make accurate predictions within the DT framework. Delving deeper into image analysis, Deep CNNs build upon CNN foundations. They, too, have found a niche in prediction analysis within DTs. This sophisticated extension of CNNs offers enhanced capabilities for processing complex visual data, contributing to more accurate predictions.

\begin{boxes}
{\boxhead{Convolutional Neural Networks (CNN)}}
{\textit{CNNs are specialized NN architectures designed primarily for processing grid-like data, such as images and videos. They employ convolutional layers to learn hierarchical features automatically.}}
\end{boxes}

Moving beyond CNN, Long Short-Term Memory (LSTM) networks are pivotal in DTs. LSTM networks, renowned for their capability to capture temporal dependencies and sequences, have been harnessed by multiple researchers. Su and Qu \cite{su2022detection}, Wu et al. \cite{wu2022deep}, and Yi et al. \cite{yi2022digital} employ LSTM to enhance security, privacy, and task optimization within DTNs. LSTM's power in understanding time-series data allows it to detect patterns and anomalies, safeguard sensitive information, and optimize task allocation. Convolutional LSTM represents a hybrid approach, combining the strengths of CNNs and LSTM networks. This amalgamation is particularly advantageous when spatiotemporal relationships must be considered. Ji et al. \cite{ji2022digital} have employed Conv-LSTM for prediction analysis within DTNs. The model's capacity to capture complex spatiotemporal dependencies makes it valuable for making precise predictions regarding future states or events.

\begin{boxes}
{\boxhead{Long Short-Term Memory (LSTM)}}
{\textit{LSTM is a Recurrent NN (RNN) architecture designed to model sequential data. LSTMs have memory cells that can capture and remember information over long sequences, making them suitable for time series analysis tasks.}}
\end{boxes}
\begin{boxes}
{\boxhead{Convolutional LSTM (C-LSTM)}}
{\textit{Convolutional Long Short-Term Memory, or C-LSTM, is a hybrid neural network architecture that combines the spatial processing capabilities of Convolutional Neural Networks (CNNs) with the sequential modeling capabilities of Long Short-Term Memory (LSTM) networks. C-LSTMs are employed in tasks involving spatiotemporal data analysis, such as video and gesture recognition.}}
\end{boxes}

Generative Adversarial Networks (GANs) have also made their presence within DTs. They are a class of models known for data generation that have been leveraged by Ferriol-Galmes et al. \cite{ferriol2022flowdt} to enhance prediction analysis within DTs. The generative abilities of GANs have been instrumental in creating synthetic data to supplement prediction tasks, ultimately improving the accuracy of predictions. Furthermore, variants of GANs, such as the Bi-directional GAN (BiGANs), with their bi-directional generative capabilities, offer a unique advantage in generative tasks and anomaly detection. Kaytaz et al. \cite{kaytaz2023graph} have successfully implemented BiGANs for fault and anomaly detection within DTs, harnessing their ability to capture complex data relationships and deviations. Another noteworthy variant is the Wasserstein GAN (WaGAN). It has been tailored by Liu et al. \cite{liu2022blockchain} to support security and privacy measures within DTNs focusing on blockchain-related security enhancements. This highlights the adaptability of GANs and their variants to address specific security challenges. The Augmented Wasserstein GAN with Gradient Penalty (AWGAN-GP) represents a specialized GAN variant designed to generate and maintain data distribution. Zhu et al. \cite{zhu2023fault} have harnessed AWGAN-GP for prediction analysis within DTs.

\begin{boxes}
{\boxhead{Generative Adversarial Networks (GAN)}}
{\textit{GANs are a class of DL models comprising two NNs, a generator, and a discriminator, which engage in a game-theoretic framework. GANs generate synthetic data by learning the underlying distribution.}}
\end{boxes}
\begin{boxes}
{\boxhead{Wasserstein Distance GAN (WaGAN)}}
{\textit{WaGAN is a variant of GANs that employs the Wasserstein distance metric to measure the dissimilarity between generated and real data distributions. WaGANs are valued for their stability and improved training in GANs.}}
\end{boxes}
\begin{boxes}
{\boxhead{Average Wasserstein GAN with Gradient Penalty (AWGAN-GP)}}
{\textit{AWGAN-GP is a variant of GANs that combines the Wasserstein GAN with a gradient penalty term in the loss function. This modification enhances training stability and encourages smoother data generation.}}
\end{boxes}

\subsection{Reinforcement Learning (RL) and Optimization Techniques}
Reinforcement Learning (RL) introduces dynamic decision-making capabilities to DTNs. These tools are characterized by their ability to make sequential decisions to optimize various aspects, including energy efficiency, latency, security, privacy, and task optimization \cite{chen2023a3c,saravanan2021performance,he2023federated,li2022digital,shui2023cell,liu2023deep,guemes2022accelerating,lu2020low,jiang2022virtual,han2022polymorphic,yi2022digital}.

Asynchronous Advantage Actor-Critic (A3C) takes the lead, focusing on enhancing energy efficiency within DTNs \cite{chen2023a3c} to optimize decision-making in environments with multiple agents. It proves invaluable in optimizing resource allocation and reducing energy consumption. Saravanan et al. \cite{saravanan2021performance} also employ A3C but specifically emphasize latency optimization. A3C's ability to adapt and learn optimal policies over time minimizes latency within DTs, ensuring real-time responsiveness and efficient data transfer.

\begin{boxes}
{\boxhead{Actor-Critic (A3C)}}
{\textit{A3C is an RL technique where two distinct components work together. The "actor" is responsible for making decisions or actions, while the "critic" evaluates those actions by estimating their value. A3C aims to optimize policies by concurrently adjusting the decision-making strategy (actor) and value estimation (critic), leading to more effective learning in complex environments.}}
\end{boxes}

Extended from RL, Deep Reinforcement Learning (DRL) encompasses a range of techniques that learn optimal policies through interactions with the environment. Liu et al. \cite{liu2023deep} employ DRL to optimize resource allocation, contributing to energy efficiency enhancements and reduced environmental impact. Another variant is the Multi-agent RL. It is a collective approach that enables collaborative learning among multiple agents, enhancing security measures and safeguarding sensitive information. In DTNs, Lu et al. \cite{lu2020low} address security and privacy concerns using multi-agent RL techniques.

\begin{boxes}
{\boxhead{Hierarchical Multi-Agent RL}}
{\textit{Hierarchical Multi-Agent RL is an approach that involves multiple agents with varying levels of decision-making authority. It introduces a hierarchical structure where high-level agents make global decisions while lower-level agents handle more detailed tasks.}}
\end{boxes}

Similarly, Q-learning has been developed as a model-free RL technique for solving Markov Decision Processes (MDPs). It estimates the value of taking specific actions in different states and iteratively updates these estimates to find an optimal policy. Q-learning is particularly useful for solving problems with uncertain or unknown environment dynamics. Deep Q-Network (DQN) is a variant of Q-learning. Shui et al. \cite{shui2023cell} concentrate on enhancing energy efficiency within DTs using DQNs. DQN's ability to approximate optimal Q-values for actions is harnessed to optimize resource allocation, reduce energy consumption, and boost sustainability in DT systems. Double DQN (DDQN) excels in learning optimal action-value functions. By applying DDQN, Li et al. \cite{li2022digital} fine-tuned latency parameters, ensuring that data processing and communication in DTs occur with minimal delay.

\begin{boxes}
{\boxhead{Double Deep Q-Network (DDQN)}}
{\textit{DDQN is a DRL. DDQN addresses overestimation issues in Q-learning by employing two separate NNs to estimate action values.}}
\end{boxes}

Proximal Policy Optimization (PPO) is an RL scheme known for its stability and efficiency in training policies. It focuses on iteratively improving policies by making small updates, ensuring that the new policy does not stay consistent with the old one. Jiang et al. \cite{jiang2022virtual} utilize PPO to fine-tune latency-related parameters, ensuring efficient task execution and data transfer within the DT ecosystem.

\begin{boxes}
{\boxhead{Multi-Agent PPO (MA-PPO)}}
{\textit{MA-PPO extends the PPO algorithm to scenarios involving multiple interacting agents. MA-PPO enables multiple agents to learn and adapt their policies in a coordinated manner, considering the influence of other agents in the environment.}}
\end{boxes}

Deep Deterministic Policy Gradient (DDPG) is an RL method suitable for continuous action spaces. In DTN, DDPG is introduced by He et al. \cite{he2023federated} to facilitate secure data sharing and analysis within federated DTNs.

\begin{boxes}
{\boxhead{Multi-Agent Deep Deterministic Policy Gradient (MADDPG)}}
{\textit{MADDPG is an extension of the Deep Deterministic Policy Gradient (DDPG) algorithm designed for multi-agent RL scenarios. MADDPG allows multiple agents to collaborate and learn in environments where their actions affect the environment and each other.}}
\end{boxes}

\subsection{Federated Learning (FL) and Collaborative Learning} 
Federated Learning is an ML approach designed for collaborative training of models across decentralized devices or edge nodes. Instead of sending raw data to a central server, FL allows devices to train models locally and share only model updates with the central server. This approach enhances data privacy and efficiency while enabling collective model improvement across a distributed network \cite{he2023federated, jiang2021cooperative, lu2020low, qian2023secrecy, qu2022fedtwin, su2022detection,sun2020dynamic}. Variants of FL schemes have been used in DTNs. Dynamic FL, exemplified by Sun et al. \cite{sun2020dynamic}, focuses on task offloading and optimizing content within DTNs. Dynamic FL's adaptive approach optimizes computational tasks and content delivery to meet specific performance requirements. The other variant, the asynchronous FL, optimizes learning by allowing participants to update their models at their own pace, minimizing data exposure. He et al. \cite{he2023federated} use asynchronous FL to enhance privacy within federated DTNs. The last variant is cooperative FL, as showcased by Jiang et al. \cite{jiang2021cooperative}. It fosters collaboration among participating entities while maintaining data isolation. This approach ensures that sensitive information remains confidential, paving the way for secure collaborative model training within DTs. Hierarchical FL (HFL) extends the concept of FL to incorporate hierarchical structures within decentralized networks. HFL introduces layers of communication and coordination, where local devices or nodes collaborate within smaller groups before sharing their updates with higher-level aggregators. This hierarchical approach enhances scalability and adaptability in large-scale federated learning systems. The FL with Secure Bi-Level Optimization (FCL-SBLS) takes a unique approach to FL by incorporating secure bi-level optimization. This method ensures that data privacy and security are paramount while allowing collaborative model training and knowledge sharing within federated DTNs \cite{he2023federated}.
\begin{boxes}
{\boxhead{Stacked Broad Learning System (FCL-SBLS)}}
{\textit{FCL-SBLS is a framework that combines the capabilities of stacked autoencoders and Broad Learning Systems. It is used for feature learning, data representation, and classification tasks. FCL-SBLS leverages FL techniques to extract hierarchical and abstract features from data, enhancing its ability to handle complex patterns.}}
\end{boxes}
\subsection{Graph and Network Analysis Techniques}
Graph-based analysis techniques have gained prominence within the context of DTNs. These approaches leverage the inherent interconnectedness of data and devices within the DT ecosystem and many challenges, such as fault and anomaly detection, latency optimization, and prediction analysis \cite{kaytaz2023graph, ferriol2022building, wang2020graph, ferriol2022flowdt, saravanan2022enabling}.

Graph Neural Networks (GNNs) are NN models that process and analyze graph-structured data. GNNs operate by propagating information through nodes and edges of a graph, enabling tasks such as node classification, link prediction, and graph-level predictions. GNNs are proficient at modeling complex relationships and dependencies in graph data. As illustrated by Kaytaz et al. \cite{kaytaz2023graph}, they find their application in fault and anomaly detection within DTs. GNNs enable the identification and mitigation of faults and anomalies, ensuring the robustness and reliability of DT systems. As demonstrated by Ferriol-Galmes et al. \cite{ferriol2022building}, by analyzing the network structure and interdependencies among components, GNNs facilitate data transfer and processing optimization, reducing latency and ensuring real-time responsiveness. Prediction analysis, a cornerstone of DTN tasks, relies on the predictive capabilities of GNNs. Wang et al. \cite{wang2020graph}, Ferriol-Galmes et al. \cite{ferriol2022flowdt}, and Saravanan et al. \cite{saravanan2022enabling} harness GNNs to make accurate predictions about future states and events. By considering the intricate relationships among data points and devices, GNNs enhance prediction accuracy, enabling informed decision-making within DT systems. Furthermore, Graph CNNs have been presented to extend the capabilities of traditional GNNs. Graph CNNs excel in modeling complex data relationships within graph structures, making them particularly suited for identifying and mitigating faults and anomalies in DT systems as employed by Kaytaz et al. \cite{kaytaz2023graph}.

\begin{boxes}
{\boxhead{Graph CNN (Graph CNN)}}
{\textit{A Graph CNN is a NN architecture specialized for processing graph data. Graph CNNs adapt convolutional operations to work on graph-structured data, allowing them to capture local and global patterns within graphs.}}
\end{boxes}

\section{ Main Challenges in AI-based DTNs}
\subsection{Key Challenges}
The application of these various AI tools within DTNs is undoubtedly promising. However, several challenges may arise in adopting and effectively utilizing these tools in the context of DTNs and their associated tasks. Some key challenges are \cite{almasan2022digital,rathore2021role,abdel2022security,shahraki2021comprehensive,ahmad2020machine,wang2021applications}:
\begin{itemize}
    \item Data Quality and Availability: AI tools heavily rely on data, and DTNs are no exception. Ensuring high-quality and diverse data is crucial for training accurate models. In DTNs, obtaining real-time, reliable, and comprehensive data can be challenging due to the complexity and dynamic nature of the mirrored physical systems. Inadequate or noisy data can lead to suboptimal model performance.
    \item Scalability: DTNs often involve vast, complex systems with many interconnected components. Scaling AI models to handle the increasing size and complexity of DTNs can be computationally intensive. A significant challenge is ensuring that AI tools can efficiently operate on large-scale DTs while maintaining real-time performance.
    \item Interpretability: Understanding and interpreting the decisions made by AI models is crucial, especially in applications like security, privacy, and fault detection within DTNs. Many advanced AI models, particularly DL and RL, are often regarded as "black boxes," making it challenging to explain their decision-making processes, which can be a barrier to trust and accountability.
    \item Robustness: The resilience of AI models to adversarial attacks and unexpected system behavior is paramount, particularly in security and privacy applications. Ensuring that AI tools can detect and adapt to novel threats or anomalies in the dynamic environment of DTNs is a significant challenge.
    \item Energy Efficiency: While AI tools can contribute to energy efficiency in DTNs, they can also be computationally intensive and power-hungry. Striking a balance between the computational demands of AI models and the energy-efficient operation of DTs is a challenge, particularly in resource-constrained environments.
    \item Privacy Concerns: Data privacy is critical in FL and collaborative settings. Ensuring that FL and other collaborative AI tools protect sensitive information while allowing for effective model training is a delicate balancing act.
    \item Latency and Real-Time Processing: Many AI tools, especially RL and FL, require iterative processes that may not align with the real-time demands of DTNs, particularly in latency-sensitive applications. Ensuring that AI tools meet real-time requirements while delivering optimal results is a significant challenge.
    \item Model Complexity: Some AI models can be highly complex and require significant computational resources. In DTNs, especially in edge and resource-constrained environments, deploying and maintaining such complex models can be challenging.
    \item Model Generalization: AI models must be able to generalize from historical data to adapt to new situations or changes in DTNs. Ensuring that models can adapt effectively when faced with previously unseen scenarios is an ongoing challenge.
    \item Integration and Interoperability: DTNs often involve a mix of heterogeneous systems, and integrating AI tools seamlessly into these environments can be complex. Ensuring compatibility, interoperability, and ease of integration with existing DT systems is crucial.
\end{itemize}

\subsection{Responsible AI Considerations}
Addressing these obstacles will foster successful and responsible AI implementation for more efficient systems. To countermeasure the challenges faced in adopting AI across its key tasks, responsible strategies must be devised for successful implementation. These strategies promote fairness, accuracy, transparency, and security \cite{jang2023digital,wang2021applications,zhang2023operationalizing}.
\begin{boxes}{\boxhead{Responsible AI}} {\textit{Responsible AI refers to the ethical and moral considerations associated with the development and deployment of AI systems. It encompasses a set of principles and practices to ensure that AI technologies are used in a way that aligns with human values and respects the rights and well-being of systems.}}
\end{boxes}
\subsubsection{Focusing on Data Quality and Representation}
\begin{itemize}
    \item \textbf{Data Preprocessing and Bias Mitigation:} Employ rigorous data preprocessing techniques to address historical biases in DTN data. Utilize diverse and representative datasets to ensure accurate predictions and avoid reliance on outdated information.
    
    \item \textbf{Enhanced Data Quality Control:} Implement robust data quality control measures to ensure reliable DT model representation. Source and integrate data from multiple reliable sources to provide a comprehensive system view. Facilitate informed decision-making and enhance network management capabilities.
    
    \item \textbf{Integration of Simulation and Real-time Data:} Integrate simulation and real-time data to improve prediction accuracy within DTNs. This integration enhances the DT's ability to make accurate predictions. Incorporate probabilistic modeling to account for unpredictable events and behaviors, resulting in more reliable performance enhancement strategies and recommendations.
\end{itemize}

\subsubsection{Focusing on AI Model Reliability and Transparency}
\begin{itemize}
    \item \textbf{Transparent AI Models:} Prioritize using explainable AI models within DTNs to enhance transparency and interpretability. Provide detailed explanations for model recommendations and strategies to build trust among stakeholders. Transparency supports informed decision-making.
    
    \item \textbf{Continuous Model Refinement:} Recognize the complexity of DTNs and the need for ongoing improvement. Continuously refine and optimize AI models to ensure their reliability in dynamic DTN environments. Validate models using real-world data to maintain effective performance.
\end{itemize}

These strategies are tailored to address the challenges and goals of DTNs. By focusing on data quality, security, privacy, and the reliability and transparency of AI models, these strategies aim to enhance various aspects of DTNs, including performance, network management, communication, prediction accuracy, anomaly detection, and overall security and privacy.

\section{Conclusion and Key Points}
\begin{boxes}
{\boxhead{DTNs Overview}}
{\begin{itemize}
        \item DTNs span a wide spectrum of physical networks, including cellular, wireless, optical, satellite, and more.
        \item These virtual networks seamlessly integrate computational power and AI capabilities, breaking conventional boundaries.
        \item Recommendations from DTNs are disseminated to physical network entities.
        \item This process operates as an autonomous closed-loop data transmission paradigm.
        \item DTNs continuously learn, evolve, and adapt based on real-world outcomes.
    \end{itemize}}
\end{boxes}

\begin{boxes}
{\boxhead{Key Tasks in DTNs}}
{\begin{itemize}
        \item Network performance enhancement, including latency optimization, energy efficiency, and Task offloading.
        \item Network management, including resource allocation planning and monitoring.
        \item Communication enhancement.
        \item Prediction analysis for traffic and KPIs.
        \item Anomaly detection, security, and privacy preservation.
    \end{itemize}}
    \end{boxes}

\begin{boxes}
{\boxhead{AI Tools in DTNs}}
{   \begin{itemize}
        \item {ML tools: NNs, Genetic algorithms, TL, and XGBoost.}
        \item {DL tools: CNNs, LSTMs, and GANs.}
        \item {RL tools: A3C, DRL, Multi-agent RL, Hierarchical RL, Q-learning, DDQN, MA-PPO, and MADDPG.}
        \item {FL tools: Dynamic FL, Asynchronous FL, Cooperative FL, and FCL-SBLS.}
        \item {Graph-Based Techniques: GNNs and Graph CNNs.}
    \end{itemize}}
    \end{boxes}

\begin{boxes}
{\boxhead{Challenges Faced by AI Schemes}}
{    \begin{itemize}
        \item Data quality and availability.
        \item Scalability.
        \item Interpretability.
        \item Robustness.
        \item Privacy concerns.
        \item Security.
        \item Energy efficiency.
        \item Model complexity.
        \item Generalization.
        \item Integration.
    \end{itemize}}
    \end{boxes}

\begin{boxes}
{\boxhead{Responsible AI Strategies}}
{    \begin{itemize}
        \item Bias Mitigation and Fairness.
        \item Transparency.
        \item Privacy preservation.
        \item Accountability.
    \end{itemize}}
\end{boxes}

In essence, DTNs represent a dynamic and evolving landscape where AI technologies, driven by domain expertise, are poised to revolutionize network performance, management, and security across many physical network domains, offering substantial benefits and opportunities for innovation.

\bibliographystyle{vancouver-modified}
\bibliography{Sample}
\end{document}